%% file: main.tex
\documentclass[article]{IEEEtran}

\input{preamble}

\newcommand{\citeauthors}[1]{\citeauthor{#1}~\cite{#1}}

\input{math}

\input{defs} %

\tikzstyle{block} = [rectangle, draw, minimum height=1.2cm, minimum width=2.8cm, text centered, text width=2.7cm, font=\small, fill=gray!10]
\tikzstyle{arrow} = [thick, -{Latex[width=2mm]}]

\begin{document}
\pagenumbering{gobble} %

\title{A Low-Cost Open-Source BLE-Based Asian Hornet Tracking System}

\author{
\IEEEauthorblockN{%
Gilles Callebaut\,\orcidlink{0000-0003-2413-986X}\IEEEauthorrefmark{1},
Jan Van Moer\IEEEauthorrefmark{2},
}\\
\IEEEauthorblockA{\IEEEauthorrefmark{1}%
KU Leuven, Belgium}\\
\IEEEauthorblockA{\IEEEauthorrefmark{2}%
Van Moer Consultancy, Belgium} %
}

\maketitle 
\defineauthors[false]{todo, gilles, jan}

\begin{abstract}

\input{sections/00_abstract}

\end{abstract}


\section{Introduction}
\begin{figure}[h]
    \centering
    \begin{subfigure}[b]{0.45\linewidth}
        \centering
    \includegraphics[width=0.78\linewidth]{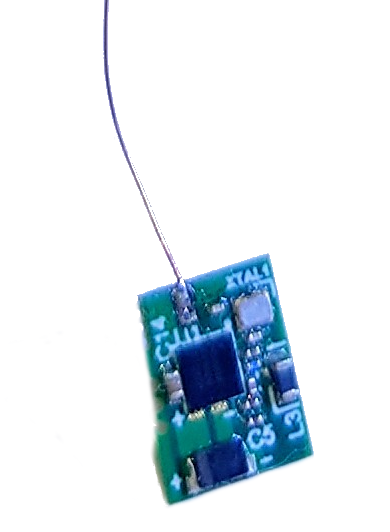}
    \caption{Tag}
    \label{fig:tag}
    \end{subfigure}~%
    \begin{subfigure}[b]{0.45\linewidth}
        \centering
        \includegraphics[width=0.9\linewidth]{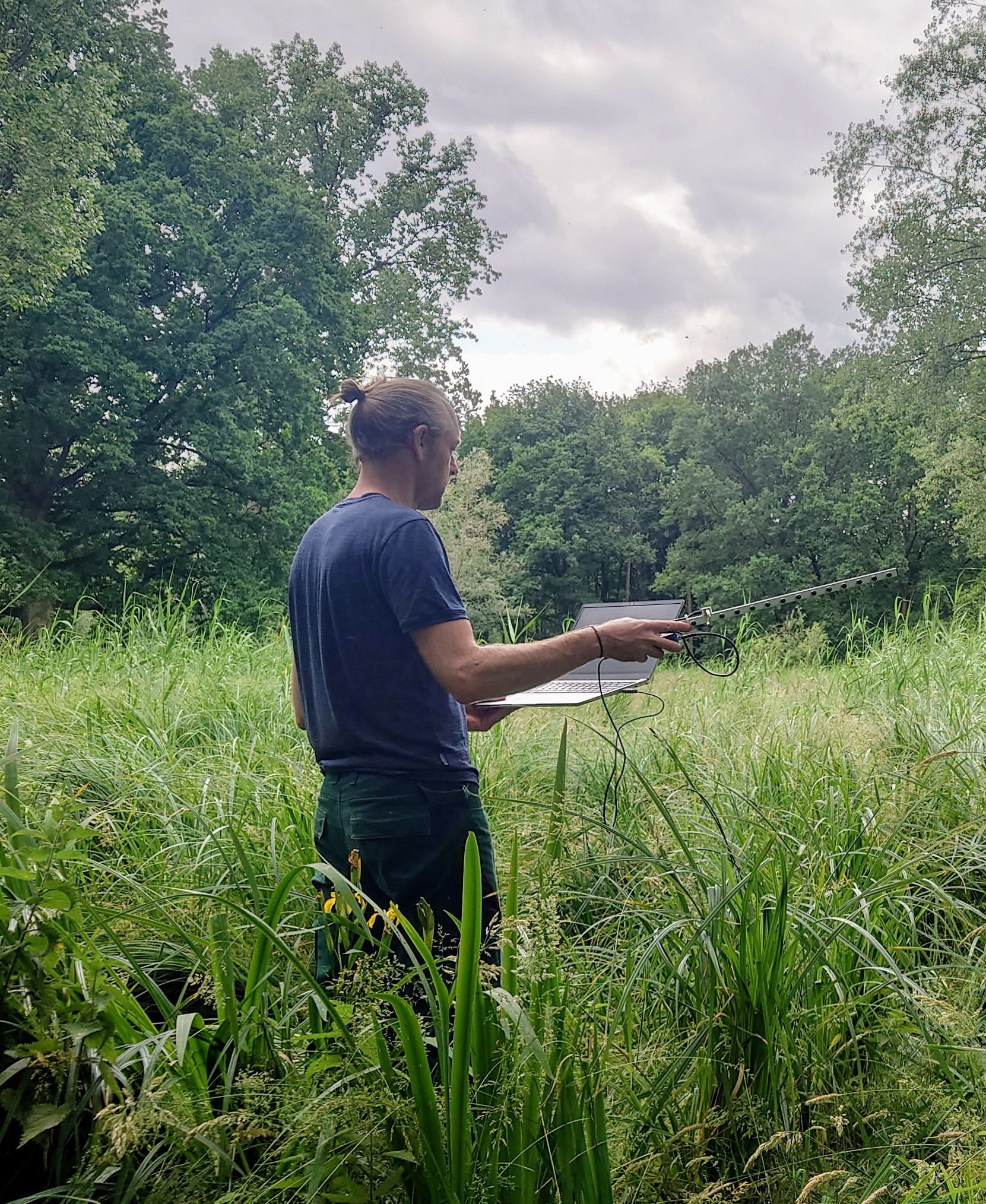}
        \caption{Reader}
    \end{subfigure}
    \caption{Images of the tag and reader.}\label{fig:setup}
\end{figure}

The emergence of the Asian hornet has led beekeepers to invest numerous hours in locating their nests. This task is primarily accomplished through the laborious method of luring hornets at three separate locations and employing triangulation techniques. However, this approach is both time-consuming and challenging to execute successfully. While there are commercial miniature transmitters available for this purpose, they tend to be prohibitively expensive and lack user-friendly features. Typically, these transmitters operate on a basic two-stage transmission system within the frequency range of \SIrange{142}{174}{MHz}. To receive transmissions, a 2-meter band receiver with a Yagi antenna is commonly utilized. The existing solutions are often cumbersome. In contrast, digital and commodity hardware offer a more practical and scalable alternative. Therefore, in this work, we employ and adapt \gls{ble} to enhance the detection range of the tracking system. The proposed approach also enables the implementation of a low-cost, fully digital beam-sweeping receiver, capable of autonomously tracking the hornet’s direction. A picture of the system, tag and reader, is shown in \cref{fig:setup}.

\textbf{System Requirements. }
The aim of the system is to track Asian hornets, in order to find and extinguish their nests. The system is required to operate with low power consumption, providing a lifetime of \num{3} to \num{6} hours when pulsing or transmitting every second. The total weight, including the battery, should be less than \SI{200}{\milli\gram}. The hornet should be able to be tracked in an area of 200 square meters. The cost per tag should be constrained to 10 to 20 euros. Finally, the time required to locate the nest is approximately \SI{1}{\hour}.

\textbf{Contributions.} In this work, we present a low-cost and open-source tracking system tailored for localizing Asian hornets using \gls{ble}. Our contributions are threefold. First, we design a lightweight BLE-based tag by directly programming the radio peripheral of an nRF52 chip, allowing for full control over the transmitted bitstream and enabling the insertion of custom \gls{pn} sequences. Second, we develop a directional SDR-based receiver in GNU Radio, equipped with a correlation block that identifies the tag's direction by matching demodulated bits with the expected PN sequence. This allows for beam sweeping and automated localization without manual antenna adjustment. Third, we validate our system through field measurements, demonstrating robust angular resolution and a maximum range of \SI{360}{\meter}. The entire system, including firmware, SDR processing chain, and measurement scripts, is released as open-source~\faGitlab~\cite{opensource}, offering a reproducible and extensible platform for the broader community working on invasive species tracking and wireless telemetry.

\section{Solutions to Tracking Hornets}

\subsection{``Wick pot'' Technique - Manual Triangularisation}

This laborious method involves luring hornets at three separate locations. It does not require advanced technology and relies on measuring the direction and time it takes for a hornet to fly from a bait site to its nest and back. By calculating this round-trip flight time, the nest's distance can be estimated through triangulation. However, this approach is both time-consuming and challenging to execute successfully.

\subsection{Imaginary}
Two prominent approaches using imagery for insect tracking involve retroreflective tags and infrared photoluminescent markers. \citeauthors{smith2021method} demonstrates the use of lightweight, inexpensive retroreflective tags that can be detected by comparing images taken with and without a flash using a global shutter camera. These tags minimally impact insect behavior and can be deployed on drones or masts, though the tracking range is limited to a few tens of meters and is less effective in forested environments. Alternatively, \citeauthors{walter2021new} explores photoluminescent tags emitting at \SI{1400}{\nano\meter}, a wavelength strongly absorbed by the atmosphere. Constructed from passivated lead sulfide quantum dots, these tags offer excellent \gls{snr} and were used to analyze bumblebee flight patterns without the interference of antenna-based tracking, with each marker weighing only \SI{12.5}{\milli\gram} and measuring \SI{5}{\milli\meter} in diameter.

Radio-telemetry approaches for animal and insect tracking span a range of technologies, each with distinct advantages and trade-offs. \gls{rfid} is widely used in sectors such as livestock management, agriculture, and wildlife conservation. As reviewed in \cite{10324350}, RFID enables monitoring of animal behavior, health, and movement patterns, with passive UHF tags being the most common due to their low power and cost. However, the range is limited, and implementations like the \SI{13}{cm} antenna wrapped around an acorn for tracking zoochoric dispersal~\cite{electronics13030567} are impractical in miniaturized setups such as insect tracking.

Harmonic radar offers another option, enabling detection of insects carrying passive transponders that reflect the second harmonic of an incident signal~\cite{maggiora2019innovative}. This allows for precise localization while rejecting background clutter. Systems can achieve ranges up to 500\,m, using either single-stage or more powerful two-stage transmitters. The single-stage variant is compact and easier to maintain but limited in power and harmonic purity~\cite{lu2021frequency}, while the two-stage approach enhances performance at the cost of complexity~\cite{naef2005miniaturization}. Both rely on high-gain directional antennas and mechanical rotation for beam steering, which can be cumbersome in the field and requires line of sight.

In contrast, Bluetooth-based approaches, particularly those leveraging Bluetooth Low Energy (BLE), are emerging as a flexible and low-cost alternative. \citeauthors{shearwood2020development} demonstrates how standard Bluetooth 4.0 chips can be repurposed by reverse-engineering the protocol to carry a lower-rate modulation scheme. This technique significantly extends communication range and allows the insertion of a virtual preamble—conceptually similar to our approach—for improved synchronization. By reducing the effective data rate to \SI{33}{\kilo\hertz}, the system enhances resilience to noise and reduces power consumption, making it highly suitable for long-range, low-power tracking scenarios such as hornet or bee localization. Unlike harmonic radar or RFID, BLE also benefits from widespread hardware availability and native digital integration, enabling the design of software-defined receivers with beam-sweeping capabilities. This eliminates the need for mechanical antenna repositioning and facilitates scalable, automated deployments using commodity components.

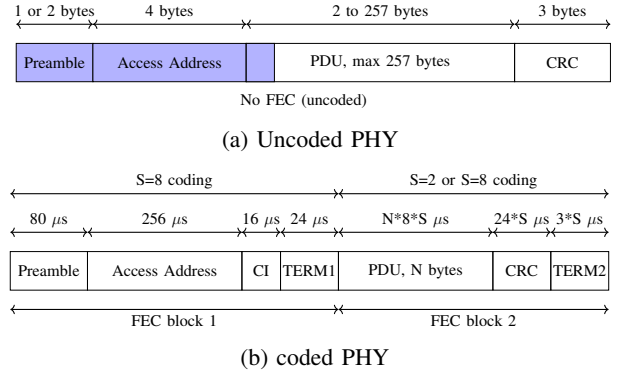
\begin{figure}[h!]
    \centering
    \begin{subfigure}[b]{0.95\linewidth}
        \centering
    \input{uncoded-phy.tex}
    \caption{Uncoded PHY}\label{fig:uncoded-phy}
    \end{subfigure}
    \vspace{0.2cm}
    
    \begin{subfigure}[b]{0.95\linewidth}
        \centering
         \input{coded-phy.tex}
    \caption{coded PHY}\label{fig:coded-phy}
    \end{subfigure}
    \caption{Link Layer packet format for the LE Uncoded and Coded PHY (not in scale). A total of \num{256}~bits is used for the \gls{pn} code, i.e.,  \num{1} byte preamble,  \num{4} bytes access address,  \num{27} bytes PDU.}
\end{figure}

\section{Our solution - Design}

The design is publicly available on GitLab~\faGitlab~\cite{opensource}.
The installation exists of a hornet tag, equipped with a Bluetooth chip, and a reader processing the incoming Bluetooth signal in GNU~Radio.

\input{gnu-radio.tex}

\subsection{Hornet Tag}
The design of the hornet tag opted for a \gls{ble} modem due to its low weight and small size, making it suitable for the application. However, one limitation of this approach is that it is restricted to the supported stack and modulation schemes of Bluetooth. To address this, the approach was to use the \gls{ble} PHY with a custom payload, designed to act as a \gls{pn} sequence. This allows us to leverage coding gain and improve receiver sensitivity. The development started with the nRF52 SDK, but without the need for the full \gls{ble} stack, as we are not implementing the \gls{ble} protocol. Instead, we directly program on top of the radio peripheral, enabling more flexibility in the design.

We have the ability to modify and program the access address and \gls{pdu}, allowing us to create a custom preamble. Additionally, the radio is configured to operate at the maximum transmit power, which for the nRF52820 is \SI{8}{dBm}. The antenna is a copper wire, measuring \SI{12.5}{cm}, which corresponds to the wavelength.

Bluetooth introduced the \textit{coded PHY} for long-range communication, which uses channel coding, i.e., \gls{fec}, to recover bit errors. The frame structure of the coded PHY is shown in \cref{fig:coded-phy}. As illustrated, for each payload bit, \num{2} (S2) or \num{8} (S8) bits are generated, which reduces the effective throughput by a factor of \num{2} and \num{8}, respectively. The coded PHY uses a preamble consisting of \num{80} symbols and includes \num{10} repetitions of the symbol pattern "\texttt{00111100}". 

To detect a packet,  we apply a correlation-based technique on the demodulated bits. This allows to cope with bit errors, and gives us the best correlation in the direction of the transmitter (see \cref{sec:reader}). To implement this, a \gls{pn} code is embedded within the frame. Hence, we want a frame structure that allows us to generate the longest \gls{pn} sequence. Given the fixed, and long preamble and \gls{fec} of the coded PHY, we utilize the \textit{uncoded PHY} frame format (\cref{fig:uncoded-phy}). The \textit{Access Address} and \textit{\gls{pdu}} are designed to generate a long unique \gls{pn} sequence. Each tag has its own unique \gls{pn} sequence, allowing different tags to be distinguishable from each other due to the low cross-correlation between different \gls{pn} sequences.

To be complete, other sequences with good autocorrelation characteristics do exist, but they are not usable in this setup due to the binary data format we can manipulate with the BLE modem. For instance, Zadoff-Chu sequences, while offering excellent autocorrelation properties, require operating on complex symbols. This makes Zadoff-Chu sequences and similar options impractical for our approach, where simplicity and efficient manipulation of binary data are essential.

\subsection{Reader}\label{sec:reader}

The reader consists of a \gls{sdr}, attached to a computer running \gls{gr}. For our experiments, a 16-element \SI{2.4}{GHz} directional \SI{16}{dBi} antenna Yagi antenna\footnote{\url{https://www.farnell.com/datasheets/1580319.pdf}} and a PlutoSDR is used. A block diagram of the different blocks implemented in \gls{gr} is depicted in \cref{fig:block-sdr}. The PlutoSDR samples at \SI{4}{MS/s}, oversampling the signal by a factor of \num{4}, as \gls{ble} uses \SI{1}{\mega\bit\per\second} modulation. A \texttt{Power Squelch} block is used to only demodulate the incoming signal when the power is above a certain configured threshold. The \gls{gmsk} symbols are then demodulated and fed to the custom-designed correlation block. This block outputs the correlation between the demodulated bits and the expected \gls{pn} sequence.

\section{Evaluation}

The proposed solution has been evaluated with respect to both its angular sensitivity and achievable communication range. A maximum distance of \SI{360}{\meter} was successfully achieved. In addition, \cref{fig:circ-meas} presents the measured correlation values as a function of the azimuth angle of the Yagi antenna. During this experiment, the tag was positioned at a fixed distance of \SI{50}{\meter}, and the correlation with respect to different antenna orientations was recorded. The correlation peaks at \SI{0}{\degree}, which corresponds to the direct line-of-sight direction toward the tag. The maximum correlation value observed is \num{256}, which corresponds to the full length of the employed \gls{pn} sequence.

\input{circ-mean.tex}

In terms of performance, our custom modulation approach achieves a slightly better range than the standard \gls{ble} coded PHY in initial tests. However, a notable drawback of this method lies in the increased computational demands on the receiver side. Despite this limitation, there is potential for further improvement. By increasing the length of the \acrlong{pn} sequence and combining correlation with power-based measurements, both the computational efficiency and the effective communication range can be enhanced.

An important advantage of our approach is its flexibility. The custom preamble can be further tailored, and techniques such as frequency hopping or spreading the transmission across multiple \gls{ble} channels can be integrated to enhance robustness. Moreover, the directional nature of the antenna system lends itself well to future beamforming strategies, enabling more precise and automated tracking capabilities.

\FloatBarrier%
\section{Conclusions and Future Work}

In this work, we give an overview of different techniques to track animals and particular small flying insects. We propose a small and low-cost solution using \gls{ble} to track Asian hornets. The goal is to locate the nests in an early stage, to eradicate the invasive species. 

Given the widespread adoption and availability of \glspl{ble} connectivity, the work can be extended to allow for community-based tracking, using OpenHayStack~\cite{Heinrich_OpenHaystack_2023} or similar concepts~\cite{Bottger_GoogleFindMyTools_2024}. Furthermore, opposing to using a directive antenna, \gls{aoa} measurements can be made using an array for automatic tracking and direction finding.

\FloatBarrier%
\printbibliography

\end{document}

%% file: preamble.tex
\usepackage[T1]{fontenc}
\usepackage{graphicx} %
\usepackage{hyperref}
\usepackage{url}
\usepackage[dvipsnames,table,xcdraw]{xcolor}
\input{auto_names} %

\usepackage[backend=biber,style=ieee]{biblatex}
\AtBeginBibliography{\footnotesize}

\usepackage{tikz}
\usetikzlibrary{angles}
\usetikzlibrary{arrows}
\usetikzlibrary{arrows.meta}
\usetikzlibrary{backgrounds}
\usetikzlibrary{calc}
\usetikzlibrary{decorations.pathmorphing}
\usetikzlibrary{fit}
\usetikzlibrary{petri}
\usetikzlibrary{plotmarks}
\usetikzlibrary{positioning}
\usetikzlibrary{quotes}
\usetikzlibrary{shapes.misc}
\usetikzlibrary {shapes.multipart}
\usetikzlibrary{spy}
\usepackage{xinttools} 
\usetikzlibrary{math}

\usepackage{pgfplots}
\pgfplotsset{compat=newest}
\usepgfplotslibrary{patchplots}
\usepgfplotslibrary{fillbetween}
\usepgfplotslibrary{units}
\usepgfplotslibrary{polar}
\usepgfplotslibrary{external}

\usepackage{siunitx}

\usepackage[xindy, nonumberlist]{glossaries}
\makenoidxglossaries%
\loadglsentries{abbr}
\usepackage[capitalize]{cleveref}

\tikzset{
    every node/.append style={font=\footnotesize},
    every label/.append style={font=\footnotesize}
}
\pgfplotsset{
    every axis legend/.append style={
        fill opacity=0.8,
        draw=none,
    },
    every axis/.append style={
        x grid style={darkgray!60},
        xmajorgrids,
        y grid style={darkgray!60},
        ymajorgrids,
    }
}

\usepackage{caption}
\usepackage{subcaption}

\usepackage{orcidlink}
\usepackage{fontawesome}
\usepackage{placeins}

%% file: abbr.tex
\newacronym{2d}{2D}{two-dimensional}
\newacronym{3d}{3D}{three-dimensional}
\newacronym{3gpp}{3GPP}{3rd generation partnership project}
\newacronym{4g}{4G}{fourth-generation}
\newacronym{5g}{5G}{fifth-generation}
\newacronym{6dof}{6DoF}{six degrees of freedom}
\newacronym{6g}{6G}{sixth-generation}
\newacronym{abp}{ABP}{authentication by personalisation}
\newacronym{ac}{AC}{alternating current}
\newacronym{aclr}{ACLR}{adjacent channel leakage ratio}
\newacronym{adc}{ADC}{analog-to-digital converter}
\newacronym{adr}{ADR}{adaptive data rate}
\newacronym{aes}{AES}{Advanced Encryption Standard}
\newacronym{afe}{AFE}{analog front end}
\newacronym{ag}{AG}{array gain}
\newacronym{agps}{A-GPS}{Assisted Global Positioning System} %
\newacronym{ai}{AI}{artificial intelligence}
\newacronym{alk}{ALK}{Alkaline}
\newacronym{am}{AM}{amplitude modulation}
\newacronym{amam}{AM/AM}{amplitude modulation to amplitude modulation}
\newacronym{amc}{AMC}{automatic modulation classification}
\newacronym{amp}{AMP}{approximate message passing}
\newacronym{ampm}{AM/PM}{amplitude modulation to phase modulation}
\newacronym{aoa}{AOA}{angle-of-arrival}
\newacronym{aod}{AOD}{angle-of-departure}
\newacronym{ap}{AP}{access point}
\newacronym{api}{API}{application program interface}
\newacronym{apt}{APT}{acoustic power transfer}
\newacronym{apu}{APU}{access point unit}
\newacronym{ar}{AR}{augmented reality}
\newacronym{arima}{ARIMA}{Auto-Regressive Integrated Moving Average}
\newacronym{arp}{ARP}{Antenna Reference Point}
\newacronym{asic}{ASIC}{application specific integrated circuit}
\newacronym{ask}{ASK}{amplitude-shift keying}
\newacronym{at}{AT}{ATtention}
\newacronym{auv}{AUV}{autonomous underwater vehicle}
\newacronym{awg}{AWG}{arbitrary waveform generator}
\newacronym{awgn}{AWGN}{additive white Gaussian noise}
\newacronym{baw}{BAW}{bulk acoustic wave}
\newacronym{bb}{BB}{base-band}
\newacronym{bc}{BC}{backscatter communication}
\newacronym{bd}{BD}{backscatter device}
\newacronym{be}{BE}{Belgium}
\newacronym{ber}{BER}{bit error rate}
\newacronym{bf}{BF}{beamforming}
\newacronym{bga}{BGA}{ball grid array}
\newacronym{bldc}{BLDC}{brushless DC}
\newacronym{bod}{BOD}{Brown-Out Detection}
\newacronym{bom}{BOM}{bill of materials}
\newacronym{bpsk}{BPSK}{binary phase-shift keying}
\newacronym{bs}{BS}{base station}
\newacronym{bu}{BU}{booster unit}
\newacronym{bw}{BW}{bandwidth}
\newacronym{ca}{CA}{carrier emitter}
\newacronym{cad}{CAD}{channel activity detection}
\newacronym{cars}{CARS}{calibration reference signal}
\newacronym{cbm}{CBM}{condition based maintenance}
\newacronym{cc}{CC}{constant current}
\newacronym{ccdf}{CCDF}{complementary cumulative distribution function}
\newacronym{ccnn}{CCNN}{circular convolutional neural network}
\newacronym{ccs}{CCS}{correlative channel sounder}
\newacronym{cdf}{CDF}{cumulative distribution function}
\newacronym{cdma}{CDMA}{code division-multiple access}
\newacronym{cdrx}{CDRX}{connected mode DRX}
\newacronym{ce}{CE}{coverage enhancement}
\newacronym{ced}{CED}{cumulative energy density}
\newacronym{cf}{CF}{cell-free}
\newacronym{cfmmimo}{CF-mMIMO}{cell-free massive MIMO}
\newacronym{cfo}{CFO}{carrier frequency offset}
\newacronym{cir}{CIR}{channel impulse response}
\newacronym{cla}{CLA}{closed-loop approach}
\newacronym{clk}{CLK}{clock}
\newacronym{cmos}{CMOS}{complementary metal oxide semiconductor}
\newacronym{cost}{COST}{commercial off-the-shelf}
\newacronym{cots}{COTS}{commercial off-the-shelf}
\newacronym{cp}{CP}{cyclic prefix}
\newacronym{cpt}{CPT}{capacitive power transfer}
\newacronym{cpu}{CPU}{central-processing unit}
\newacronym{cpw}{CPW}{coplanar waveguide}
\newacronym{cqi}{CQI}{channel quality indicator}
\newacronym{cr}{CR}{coding rate}
\newacronym{crc}{CRC}{cyclic redundancy check}
\newacronym{crlb}{CRLB}{Cram\'er-Rao lower bound}
\newacronym{crs}{CRS}{cell reference signal}
\newacronym{cs}{CS}{compressed sensing}
\newacronym{cse}{CSE}{channel state estimation}
\newacronym{csi}{CSI}{channel state information}
\newacronym{csp}{CSP}{contact service point}
\newacronym{css}{CSS}{chirp spread spectrum}
\newacronym{cu}{CU}{central unit}
\newacronym{cv}{CV}{constant voltage}
\newacronym{cw}{CW}{continuous wave}
\newacronym{d2d}{D2D}{device-to-device}
\newacronym{dab}{DAB}{Digital Audio Broadcasting}
\newacronym{dac}{DAC}{digital-to-analog converter}
\newacronym{daq}{DAQ}{data acquisition system}
\newacronym{das}{DAS}{distributed antenna systems}
\newacronym{dbpsk}{DBPSK}{Differential Binary Phase Shift Keying}
\newacronym{dc}{DC}{direct current}
\newacronym{dcc}{DCC}{dynamic cooperation clustering}
\newacronym{ddc}{DDC}{digital down conversion}
\newacronym{de}{DE}{drain efficiency}
\newacronym{dft}{DFT}{discrete Fourier transform}
\newacronym{dl}{DL}{downlink}
\newacronym{dlc}{DLC}{Distributed Laser Charging}
\newacronym{dli}{DLI}{direct link interference}
\newacronym{dlt}{DLT}{Distributed Ledger Technology}
\newacronym{dm}{DM}{diffuse multipath}
\newacronym{dma}{DMA}{Direct Memory Access}
\newacronym{dmac}{DMAC}{Direct Memory Access Controller}
\newacronym{dmc}{DMC}{diffuse multipath component}
\newacronym{dmimo}{D-MIMO}{distributed MIMO}
\newacronym{doa}{DOA}{direction-of-arrival}
\newacronym{dp}{DP}{differencial privacy}
\newacronym{dpd}{DPD}{digital pre-distortion}
\newacronym{dpdk}{DPDK}{Data Plane Development Kit}
\newacronym{dram}{DRAM}{dynamic random-access memory}
\newacronym{drcs}{$\Delta$RCS}{differential-radar cross section}
\newacronym{drx}{DRX}{Discontinuous Reception Mode}
\newacronym{dsb}{DSB}{double-sideband}
\newacronym{dsl}{DSL}{digital subscriber line}
\newacronym{dsp}{DSP}{digital signal processing}
\newacronym{dss}{DSS}{Dataset Storage Standard}
\newacronym{duc}{DUC}{digital up-converter}
\newacronym{dvb}{DVB}{Digital Video Broadcasting}
\newacronym{e2e}{E2E}{end-to-end}
\newacronym{easa}{EASA}{European Union Aviation Safety Agency}
\newacronym{ebg}{EBG}{electromagnetic bandgap}
\newacronym{ec}{EC}{European Commission}
\newacronym{ecc}{ECC}{elliptic curve cryptography}
\newacronym{ecdf}{eCDF}{empirical cumulative distribution function}
\newacronym{ecdlp}{ECDLP}{elliptic curve discrete logarithm problem}
\newacronym{ecsp}{ECSP}{edge computing service point}
\newacronym{edlc}{EDLC}{electrostatic double-layer capacitors}
\newacronym{edrx}{eDRX}{Extended Discontinuous Reception Mode}
\newacronym{ee}{EE}{energy efficiency}
\newacronym{egprs}{EGPRS}{Enhanced Data Rates for GSM Evolution}
\newacronym{eh}{EH}{energy harvesting}
\newacronym{eirp}{EIRP}{equivalent isotropically radiated power}
\newacronym{em}{EM}{electromagnetic}
\newacronym{embb}{eMBB}{enhanced Mobile Broadband}
\newacronym{en}{EN}{energy neutral}
\newacronym{end}{END}{energy-neutral device}
\newacronym{enob}{ENOB}{effective number of bits}
\newacronym{eol}{EoL}{end of life}
\newacronym{epu}{EPU}{edge processing unit}
\newacronym{er}{ER}{Energy Receiver}
\newacronym{erp}{ERP}{effective radiation power}
\newacronym[plural=ESCs,firstplural=electronic speed controllers (ESCs)]{esc}{ESC}{electronic speed control}
\newacronym{esd}{ESD}{electrostatic discharge}
\newacronym{esl}{ESL}{electronic shelf label}
\newacronym{esr}{ESR}{equivalent series resistance}
\newacronym{et}{ET}{Energy Transmitter}
\newacronym{etsi}{ETSI}{European Telecommunications Standards Institute}
\newacronym{evd}{EVD}{eigenvalue decomposition}
\newacronym{evm}{EVM}{Error Vector Magnitude}
\newacronym{ewlb}{eWLB}{Embedded Wafer Level Ball Grid Array}
\newacronym{fa}{FA}{federation anchor}
\newacronym{fair}{FAIR}{Findability, Accessibility, Interoperability, and Reuse of digital assets}
\newacronym{fd}{FD}{front-haul distance}
\newacronym{fde}{FDE}{frequency domain equalizer}
\newacronym{fdm}{FDM}{frequency-division multiplexing}
\newacronym{fdma}{FDMA}{frequency division multiple access}
\newacronym{fem}{FEM}{finite element analysis}
\newacronym{fembb}{feMBB}{further enhanced mobile broadband}
\newacronym{fft}{FFT}{fast Fourier transform}
\newacronym{fh}{FH}{fronthaul}
\newacronym{fhss}{FHSS}{frequency hopping spread spectrum}
\newacronym{fifo}{FIFO}{First In, First Out}
\newacronym{fim}{FIM}{Fisher information matrix}
\newacronym{fits}{FITS}{Flexible Image Transport System}
\newacronym{fl}{FL}{federated learning}
\newacronym{fom}{FoM}{figure of merit}
\newacronym{fpga}{FPGA}{field-programmable gate array}
\newacronym{fram}{FRAM}{Ferroelectric Random Access Memory}
\newacronym{fsk}{FSK}{frequency shift keying}
\newacronym{fss}{FSS}{frequency selective surface}
\newacronym{gb}{GB}{grant-based}
\newacronym{gdpr}{GDPR}{general data protection regulation}
\newacronym{gf}{GF}{grant-free}
\newacronym{gmsk}{GMSK}{Gaussian minimum-shift keying}
\newacronym{gnb}{gNB}{Next Generation Node B}
\newacronym{gni}{GNI}{gross national income}
\newacronym{gnn}{GNN}{graph neural network}
\newacronym{gnss}{GNSS}{global navigation satellite system}
\newacronym{gpclk}{GPCLK}{general purpose clock}
\newacronym{gpio}{GPIO}{General-Purpose Input/Output}
\newacronym{gpl}{GPL}{GNU General Public License}
\newacronym{gprs}{GPRS}{General Packet Radio Services}
\newacronym{gps}{GPS}{Global Positioning System}
\newacronym{gpu}{GPU}{graphical processing unit}
\newacronym{grc}{GRC}{GNU Radio Companion}
\newacronym{gscm}{GSCM}{geometry‐based stochastic model}
\newacronym{gsm}{GSM}{Global System for Mobile Communications}  %
\newacronym{gwp}{GWP}{Global Warming Potential}
\newacronym{harq}{HARQ}{hybrid automatic repeat request}
\newacronym{hat}{HAT}{hardware attached on top}
\newacronym{hcs}{HCS}{human-centric services}
\newacronym{hdf5}{HDF5}{Hierarchical Data Format version 5}
\newacronym{hfss}{HFSS}{High Frequency Simulator Software}
\newacronym{hmd}{HMD}{head-mounted display}
\newacronym{hpbm}{HPBM}{half power beam width}
\newacronym{HyMPRo}{HyMPRo}{Hybrid Multi-Path Routing algorithm}
\newacronym{i.i.d.}{i.i.d.}{independent and identically distributed}
\newacronym{i2c}{I2C}{Inter-Integrated Circuit}
\newacronym{iaq}{IAQ}{Indoor Air Quality}
\newacronym{ib}{IB}{in-band}
\newacronym{ibo}{IBO}{input back-off}
\newacronym{ic}{IC}{integrated circuit}
\newacronym{ici}{ICI}{intercarrier interference}
\newacronym{icnirp}{ICNIRP}{International Commission on Non-Ionizing Radiation Protection}
\newacronym{id}{ID}{information decoding}
\newacronym{idft}{IDFT}{inverse discrete Fourier transform}
\newacronym{if}{IF}{intermediate-frequency}
\newacronym{iid}{i.i.d.}{independently and identically distributed}
\newacronym{iis}{IIS}{integrated information system}
\newacronym{im}{IM}{intermodulation}
\newacronym{imd}{IMD}{intermodulation distortion}
\newacronym{imu}{IMU}{inertial measurement unit}
\newacronym{io}{IO}{input/output}
\newacronym{ioe}{IoE}{Internet of Everything}
\newacronym{iot}{IoT}{Internet of Things}
\newacronym{ipt}{IPT}{inductive power transfer}
\newacronym{ipy}{IPY}{Interventions per Year}
\newacronym{iq}{IQ}{in-phase and quadrature}
\newacronym{iqi}{IQI}{IQ imbalance}
\newacronym{ir}{IR}{infrared}
\newacronym{isi}{ISI}{intersymbol interference}
\newacronym{ism}{ISM}{industrial, scientific and medical}
\newacronym{isp}{ISP}{internet service provider}
\newacronym{jesd}{JESD}{Joint Electron Devices Engineering Council}
\newacronym{jfet}{JFET}{junction field effect transistor}
\newacronym{kpi}{KPI}{key performance indicator}
\newacronym{kvi}{KVI}{key value indicator}
\newacronym{larva}{LARVA}{LARge Virtual Array}
\newacronym{lca}{LCA}{life cycle assessment}
\newacronym{lco}{LCO}{lithium cobalt oxide}
\newacronym{ldo}{LDO}{Low-dropout voltage regulator}
\newacronym{ldpc}{LDPC}{low-density parity-check}
\newacronym{led}{LED}{Light Emitting Diode}
\newacronym{less}{LESS}{Low Energy Scheduler Solution}
\newacronym{lfp}{LFP}{lithium iron phosphate}
\newacronym{lib}{LIB}{Lithium-Ion Battery}
\newacronym{lic}{LIC}{lithium-ion capacitor}
\newacronym{lid}{LID}{Lithium Iron Disulfide}
\newacronym{lidar}{LiDAR}{light detection and ranging}
\newacronym{liion}{Li-ion}{lithium-ion}
\newacronym{lipo}{LiPo}{lithium polymer}
\newacronym{lis}{LIS}{large intelligent surface}
\newacronym{llh}{LLH}{log-likelihood}
\newacronym{lmd}{LMD}{Lithium Manganese Dioxide}
\newacronym{lmmse}{LMMSE}{least minimum mean square error}
\newacronym{lmo}{LMO}{lithium ion manganese oxide}
\newacronym{lna}{LNA}{low-noise amplifier}
\newacronym{lo}{LO}{local oscillator}
\newacronym{lora}{LoRa}{long range}
\newacronym{lorawan}{LoRaWAN}{long-range wide-area network}
\newacronym{los}{LoS}{line-of-sight}
\newacronym{lp}{LP}{linear programming}
\newacronym{lpf}{LPF}{low-pass filter}
\newacronym{lpt}{LPT}{laser power transfer}
\newacronym{lpwa}{LPWA}{Low Power Wide Area}
\newacronym{lpwan}{LPWAN}{low-power wide-area network}
\newacronym{lpwans}{LPWANs}{Low-Power Wide-Area Networks}
\newacronym{lqi}{LQI}{link quality indicator}
\newacronym{lrelu}{LReLU}{leaky rectified linear unit}
\newacronym{lrt}{LRT}{likelihood-ratio test}
\newacronym{ls}{LS}{least squares}
\newacronym{lsa}{LSA}{large synthetic array}
\newacronym{lsf}{LSF}{large-scale fading}
\newacronym{lsfc}{LSFC}{large-scale fading component}
\newacronym{lstm}{LSTM}{Long Short-Term Memory}
\newacronym{ltc}{LTC}{lithium thionyl chloride}
\newacronym{lte}{LTE}{Long Term Evolution}
\newacronym{lti}{LTI}{linear time-invariant}
\newacronym{lto}{LTO}{lithium titanate}
\newacronym{lusta}{LUSTA 5G }{Logistique mUltimodale Sécuritaire Téléopérée \& Autonome 5G}
\newacronym{m2m}{M2M}{machine to machine}
\newacronym{mac}{MAC}{Medium Access Control}
\newacronym{mate}{MATE}{millimeter-wave MIMO testbed}
\newacronym{mc}{MC}{Monte Carlo}
\newacronym{mcl}{MCL}{Maximum Coupling Loss}
\newacronym{mcs}{MCS}{modulation and coding scheme}
\newacronym{mcu}{MCU}{microcontroller unit}
\newacronym{mec}{MEC}{multi-access edge computing}
\newacronym{mems}{MEMS}{micro-electromechanical systems}
\newacronym{mf}{MF}{matched filter}
\newacronym{mimo}{MIMO}{multiple-input multiple-output}
\newacronym{miso}{MISO}{multiple-input single-output}
\newacronym{ml}{ML}{machine learning}
\newacronym{mlp}{MLP}{multilayer perceptron}
\newacronym{mmic}{MMIC}{monolithic microwave integrated circuit}
\newacronym{mmimo}{mMIMO}{massive MIMO}
\newacronym{mmse}{MMSE}{minimum mean square error}
\newacronym{mmtc}{mMTC}{massive machine-typed communication}
\newacronym{mmwave}{mmWave}{millimeter wave}
\newacronym{mn}{MN}{matching network}
\newacronym{mosfet}{MOSFET}{metal-oxide semiconductor field effect transistor}
\newacronym{mpc}{MPC}{multipath component}
\newacronym{mppt}{MPPT}{maximum power point tracking}
\newacronym{mr}{MR}{maximum ratio}
\newacronym{mrc}{MRC}{maximum ratio combining}
\newacronym{mrc_em}{MRC}{maximum ratio combining}
\newacronym{mrc_EM}{MRC}{Magnetic Resonance Coupling}
\newacronym{mrt}{MRT}{maximum ratio transmission}
\newacronym{mse}{MSE}{mean square error}
\newacronym{msk}{MSK}{Minimum-Shift Keying}
\newacronym{mtc}{MTC}{Machine-Type Communication}
\newacronym{multi-rat}{Multi-RAT}{multiple radio access technology}
\newacronym{multirat}{Multi-RAT}{Multiple Radio Access Technology}
\newacronym{music}{MUSIC}{MUltiple SIgnal Classification}
\newacronym{navauwall}{NAVAUWALL}{AUtomated NAVigation in WALLonia}
\newacronym{nb}{NB}{narrowband}
\newacronym{nbiot}{NB-IoT}{narrowband IoT}
\newacronym{nca}{NCA}{nickel cobalt aluminum}
\newacronym{netcdf}{NetCDF}{Network Common Data Form}
\newacronym{nf}{NF}{noise figure}
\newacronym{nfc}{NFC}{near-field communication}
\newacronym{nfv}{NFV}{network function virtualization}
\newacronym{ngmn}{NGMN}{Next Generation Mobile Networks }
\newacronym{ni}{NI}{National Instruments}
\newacronym{nicd}{NiCd}{nikkel cadmium}
\newacronym{nimh}{NiMH}{nikkel metal hydride}
\newacronym{nlos}{NLoS}{non-line-of-sight}
\newacronym{nmc}{NMC}{nickel manganese cobalt}
\newacronym{nmos}{nMOS}{n-channel metal-oxide semiconductor}
\newacronym{nn}{NN}{neural network}
\newacronym{nnls}{NNLS}{non-negative least squares}
\newacronym{noma}{NOMA}{non-orthogonal multiple access}
\newacronym{np}{NP}{Neyman-Pearson}
\newacronym{npbch}{NPBCH}{Narrowband Physical Broadcast Channel}
\newacronym{npss}{NPSS}{Narrow Band Primay Synchronization Signal}
\newacronym{nr}{NR}{New Radio}
\newacronym{nrs}{NRS}{Narrow Band Reference Signal}
\newacronym{nsss}{NSSS}{Narrowband Secondary Synchronization Signal}
\newacronym{ntp}{NTP}{network time protocol}
\newacronym{oai}{OAI}{OpenAirInterface} %
\newacronym{obw}{OBW}{occupied bandwidth}
\newacronym{ofdm}{OFDM}{orthogonal frequency-division multiplexing}
\newacronym{ofdma}{OFDMA}{orthogonal frequency-division multiple access}
\newacronym{ofdmim}{OFDM-IM}{OFDM with index modulation}
\newacronym{oma}{OMA}{orthogonal multiple access}
\newacronym{oob}{OOB}{out-of-band}
\newacronym{ook}{OOK}{on-off keying}
\newacronym{oran}{O-RAN}{open radio-access network}
\newacronym{os}{OS}{operating system}
\newacronym{ota}{OTA}{over-the-air}
\newacronym{otaa}{OTAA}{over-the-air authentication}
\newacronym{p1}{P1}{Phase 1}
\newacronym{p2}{P2}{Phase 2}
\newacronym{p2p}{P2P}{point-to-point}
\newacronym{pa}{PA}{power amplifier}
\newacronym{pae}{PAE}{power-added efficiency}
\newacronym{pana}{PanA}{Panel A}
\newacronym{panb}{PanB}{Panel B}
\newacronym{papr}{PAPR}{peak-to-average power ratio}
\newacronym{pc}{PC}{pilot count}
\newacronym{pcb}{PCB}{printed circuit board}
\newacronym{pcg}{PCG}{power consumption gain}
\newacronym{pcie}{PCIe}{Peripheral Component Interconnect Express}
\newacronym{pcsi}{PCSI}{perfect channel state information}
\newacronym{pd}{PD}{powered device}
\newacronym{pdcch}{PDCCH}{physical downlink control channel}
\newacronym{pdf}{PDF}{probability density function}
\newacronym{pdp}{PDP}{power delay profile}
\newacronym{pdsch}{PDSCH}{physical downlink shared channel}
\newacronym{pe}{PE}{processing element}
\newacronym{peb}{PEB}{positioning error bound}
\newacronym{per}{PER}{packet error rate}
\newacronym{pet}{PET}{privacy enhancing technology}
\newacronym{pg}{PG}{path gain}
\newacronym{pgd}{PGD}{proximal gradient descent}
\newacronym{phy}{PHY}{physical}
\newacronym{pki}{PKI}{public key infrastucture}
\newacronym{pl}{PL}{path loss}
\newacronym{pll}{PLL}{phase-locked loop}
\newacronym{pmf}{PMF}{polymer microwave fiber}
\newacronym{pmu}{PMU}{Power Management Unit}
\newacronym{pn}{PN}{pseudo-noise}
\newacronym{poe}{PoE}{power-over-Ethernet}
\newacronym{pps}{1PPS}{pulse per second}
\newacronym{pr}{PR}{phase reversal}
\newacronym{prbs}{PRBs}{Physical Resource Blocks}
\newacronym{prs}{PRS}{Peripheral Reflex System}
\newacronym{ps}{PS}{Processing System}
\newacronym{psd}{PSD}{power spectral density}
\newacronym{pse}{PSE}{power sourcing equipment}
\newacronym{psk}{PSK}{phase shift keying}
\newacronym{psm}{PSM}{power saving mode}
\newacronym{pss}{PSS}{primary synchronisation signal}
\newacronym{ptp}{PTP}{precision-time protocol}
\newacronym{ptrs}{PTRS}{Phase-Tracking Reference Signals}
\newacronym{ptw}{PTW}{paging time window}
\newacronym{pv}{PV}{photovoltaic}
\newacronym{pw}{PW}{planar wavefront}
\newacronym{pwm}{PWM}{pulse width modulation}
\newacronym{qam}{QAM}{quadrature amplitude modulation}
\newacronym{qos}{QoS}{quality-of-service}
\newacronym{qpsk}{QPSK}{quadrature phase-shift keying}
\newacronym{qrrls}{QR-RLS}{QR decomposition based recursive least squares}
\newacronym{quadriga}{QuaDRiGa}{QUAsi Deterministic RadIo channel GenerAtor}
\newacronym{ra}{RA}{Random Access}
\newacronym{ram}{RAM}{random-access memory}
\newacronym{ran}{RAN}{radio access network}
\newacronym{rar}{RAR}{Random Access Response}
\newacronym{rat}{RAT}{radio access technology}
\newacronym{rb}{Rb}{Rubidium}
\newacronym{ru}{RU}{radio unit}
\newacronym{rbs}{RBS}{radio base station}
\newacronym{rbw}{RBW}{resolution bandwidth}
\newacronym{rcs}{RCS}{radar cross section}
\newacronym{rdl}{RDL}{redistribution layer}
\newacronym{re}{RE}{radio element}
\newacronym{relu}{ReLU}{rectified linear unit}
\newacronym{rf}{RF}{radio frequency}
\newacronym{rfeh}{RFEH}{radio frequency energy harvesting}
\newacronym{rfic}{RFIC}{radio-frequency integrated circuit}
\newacronym{rfid}{RFID}{radio frequency identification}
\newacronym{rfpt}{RFPT}{radio frequency power transfer}
\newacronym{rfsoc}{RFSoC}{Radio Frequency System-on-Chip}
\newacronym{rir}{RIR}{room impulse response}
\newacronym{ris}{RIS}{reflective intelligent surface}%
\newacronym{rllmtc}{RLLMTC}{reliable low latency machine type communication}
\newacronym{rls}{RLS}{recursive least squares}
\newacronym{rms}{RMS}{root-mean-square}
\newacronym{rmse}{RMSE}{root-mean-square error}
\newacronym{rmt}{RMT}{random matrix theory}
\newacronym{rof}{RoF}{radio-over-fiber}
\newacronym{ros}{ROS}{robot operating system}
\newacronym{rpi}{RPi}{Raspberry Pi}
\newacronym{rrc}{RRC}{Radio Resource Connection}
\newacronym{rreq}{RREQ}{route request packet}
\newacronym{rsrp}{RSRP}{Reference Signals Received Power}
\newacronym{rsrq}{RSRQ}{Reference Signal Received Quality}
\newacronym{rss}{RSS}{received signal strength}
\newacronym{rssi}{RSSI}{received signal strength indicator}
\newacronym{rtc}{RTC}{real time clock}
\newacronym{rtf}{RTF}{reader talks first}
\newacronym{rtk}{RTK}{real time kinematics}
\newacronym{rts}{RTS}{ray tracing simulator}
\newacronym{rv}{RV}{random variable}
\newacronym{rw}{RW}{RadioWeaves}
\newacronym{rx}{RX}{receiver}
\newacronym{rzf}{RZF}{regularized zero forcing}
\newacronym{s-parameter}{S-parameter}{scattering parameter}
\newacronym{sa}{SA}{synchronization anchor}
\newacronym{sa5}{SA}{Stand Alone}
\newacronym{sar}{SAR}{specific absorption rate}
\newacronym{sbl}{SBL}{sparse Bayesian learning}
\newacronym{scfdma}{SCFDMA}{single-carrier frequency division multiple access}
\newacronym{sdg}{SDG}{Sustainable Development Goal}
\newacronym{sdm}{SDM}{sigma-delta modulator}
\newacronym{sdma}{SDMA}{spatial-division multiple access}
\newacronym{sdn}{SDN}{software-defined network}
\newacronym{sdof}{SDoF}{sigma-delta over fiber}
\newacronym{sdr}{SDR}{software-defined radio}
\newacronym{se}{SE}{spectral efficiency}
\newacronym{sei}{SEI}{specific emitter identification}
\newacronym{ser}{SER}{symbol-error rate}
\newacronym{sf}{SF}{spreading factor}
\newacronym{sfn}{SFN}{single frequency network}
\newacronym{sfp}{SFP}{small form-factor pluggable}
\newacronym{sha}{SHA}{Secure Hash Algorithm}
\newacronym{SigMF}{SigMF}{Signal Metadata Format}
\newacronym{simo}{SIMO}{single-input multiple-output}
\newacronym{sinr}{SINR}{signal-to-interference-plus-noise ratio}
\newacronym{siso}{SISO}{single-input single-output}
\newacronym{slam}{SLAM}{simultaneous localization and mapping}
\newacronym{slc}{SLC}{spatial leakage suppression}
\newacronym{slerp}{SLERP}{spherical linear interpolation}
\newacronym{sma}{SMA}{SubMiniature version A}
\newacronym{smc}{SMC}{specular multipath component}
\newacronym{smps}{SMPS}{switched mode power supply}
\newacronym{sndr}{SNDR}{signal-to-noise-and-distortion ratio}
\newacronym{snidr}{SNIDR}{signal-to-noise-and-interference-and-distortion ratio}
\newacronym{snir}{SNIR}{signal-to-interference-plus-noise ratio}
\newacronym{snr}{SNR}{signal-to-noise ratio}
\newacronym{soc}{SoC}{state of charge}
\newacronym{SoC}{SOC}{System on Chip}
\newacronym{sp}{SP}{service point}
\newacronym{spdt}{SPDT}{single pole double throw}
\newacronym{spi}{SPI}{Serial Peripheral Interface}
\newacronym{spst}{SPST}{single pole single throw}
\newacronym{sram}{SRAM}{static random-access memory}
\newacronym{srd}{SRD}{short-range device}
\newacronym{srls}{SRLS}{standard recursive least squares}
\newacronym{srs}{SRS}{Sounding Reference Signal}
\newacronym{ssb}{SSB}{synchronisation signal block}
\newacronym{ssd}{SSD}{solid state drive}
\newacronym{ssq}{SSQ}{simulator sickness questionnaire}
\newacronym{steam}{STEAM}{science, technology, engineering, the arts, and mathematics}
\newacronym{svd}{SVD}{singular value decomposition}
\newacronym{sw}{SW}{spherical wavefront}
\newacronym{swipt}{SWIPT}{simultaneous wireless information and power transfer}
\newacronym{synce}{SyncE}{Synchronous Ethernet}
\newacronym{tau}{TAU}{tracking area update}
\newacronym{tcer}{TCER}{transported to consumed energy ratio}
\newacronym{tcp}{TCP}{Transmission Control Protocol}
\newacronym{tcxo}{TCXO}{temperature compensated crystal oscillator}
\newacronym{tdd}{TDD}{time division duplexing}
\newacronym{tdma}{TDMA}{time division-multiple access}
\newacronym{tdoa}{TDOA}{time-difference-of-arrival}
\newacronym{thz}{THz}{Terahertz}
\newacronym{to}{TO}{timing offset}
\newacronym{toa}{TOA}{time-of-arrival}
\newacronym{tof}{ToF}{time-of-flight}
\newacronym{tosm}{TOSM}{through-open-short-match}
\newacronym{tpms}{TPMS}{Tire-Pressure Monitoring System}
\newacronym{trl}{TRL}{technology readyness level}
\newacronym{trp}{TRP}{Transmission Reception Point}
\newacronym{tsn}{TSN}{time-sensitive networking}
\newacronym{ttf}{TTF}{tag talks first}
\newacronym{ttff}{TTFF}{Time To First Fix}
\newacronym{ttm}{TTM}{time to market}
\newacronym{ttn}{TTN}{The Things Network}
\newacronym{tx}{TX}{transmitter}
\newacronym{uart}{UART}{Universal Asynchronous Receiver/Transmitter}
\newacronym{uav}{UAV}{unmanned aerial vehicle}
\newacronym{uc}{UC}{use case}
\newacronym{ucie}{UCIe}{Universal Chiplet Interconnect Express}
\newacronym{udp}{UDP}{User Datagram Protocol}
\newacronym{ue}{UE}{user equipment}
\newacronym{ugv}{UGV}{unmanned ground vehicle}
\newacronym{uhd}{UHD}{USRP hardware driver}
\newacronym{uhf}{UHF}{ultra-high frequency}
\newacronym{ul}{UL}{uplink}
\newacronym{ula}{ULA}{uniform linear array}
\newacronym{uMIMO}{$\mu$-MIMO}{ultra-massive Multiple-Input Multiple-Output}
\newacronym{ummtc}{umMTC}{ultra massive machine type communication}
\newacronym{un}{UN}{United Nations}
\newacronym{upa}{UPA}{uniform planar array}
\newacronym{ura}{URA}{uniform rectangular array}
\newacronym{urllc}{URLLC}{ultra-reliable low-latency communications}
\newacronym{usrp}{USRP}{universal software radio peripheral}
\newacronym{uv}{UV}{unmanned vehicle}
\newacronym{uwb}{UWB}{ultrawideband}
\newacronym{v2v}{V2V}{vehicle-to-vehicle}
\newacronym{vco}{VCO}{voltage-controlled oscillator}
\newacronym{vep}{VEP}{virtual edge platform}
\newacronym{vlc}{VLC}{visible light communication}
\newacronym{vlp}{VLP}{visible light positioning}
\newacronym{vna}{VNA}{vector network analyzer}
\newacronym{voc}{VOC}{Voltatile Organic Compound}
\newacronym{votable}{VOTable}{Virtual Observatory Table}
\newacronym{vr}{VR}{virtual reality}
\newacronym{wb}{WB}{wideband}
\newacronym{wban}{WBAN}{wireless body area network}
\newacronym{wd}{WD}{Wireless distance}
\newacronym{wimax}{WiMAX}{Worldwide Interoperability for Microwave Access}
\newacronym{wlan}{WLAN}{wireless LAN}
\newacronym{wpt}{WPT}{wireless power transfer}
\newacronym{wr}{WR}{White Rabbit}
\newacronym{wrsn}{WRSN}{wireless rechargeable sensor network}
\newacronym{wsn}{WSN}{wireless sensor network}
\newacronym{xets}{XETS}{cross exponentially tapered slot}
\newacronym{xr}{XR}{extended reality}
\newacronym{z3ro}{Z3RO}{zero third-order distortion}
\newacronym{zf}{ZF}{zero-forcing}
\newacronym{zmcscg}{ZMCSCG}{zero mean circularly symmetric complex Gaussian}
\newacronym{ep}{EP}{energy profiler}

\newacronym{ble}{BLE}{Bluetooth Low Energy}
\newacronym{pdu}{PDU}{Protocol Data Unit}
\newacronym{fec}{FEC}{forward-error correction}
\newacronym{gr}{GR}{GNU~Radio}

%% file: math.tex
\usepackage{amsmath,amssymb,amsfonts}
\usepackage{algorithmic}

\usepackage{xifthen}
\newcommand{\prob}[1][]{%
\ifthenelse{\isempty{#1}}%
      {\ensuremath{P}}%
    {\ensuremath{P\left\(#1\right\)}}%
}

\usepackage{amsthm}
\theoremstyle{remark}

%% file: defs.tex
\usepackage{xspace}

%% file: sections/00_abstract.tex
The Asian hornet (\textit{Vespa velutina}) poses a serious threat to ecosystems and beekeeping. Locating nests is essential, but usually involves time-consuming manual triangulation. We present a low-cost, open-source tracking system based on \gls{ble}. The system consists of a lightweight BLE tag and a \gls{sdr} receiver implemented in GNU Radio. By bypassing the BLE stack, we embed a custom \gls{pn} sequence in the uncoded PHY for correlation-based detection. Using a Yagi antenna and PlutoSDR, the receiver performs digital beam sweeping to determine the tag's direction. Field tests show reliable angular resolution at \SI{50}{\meter} and a communication range up to \SI{360}{\meter}. While our modulation increases receiver complexity, it enables future improvements such as multichannel spreading and tag identification. The design is fully open-source and provides a scalable framework for hornet tracking and related applications in environmental monitoring.

%% file: uncoded-phy.tex
\resizebox{0.95\linewidth}{!}{%
\begin{tikzpicture}
\draw[fill=blue!30] (0,0) rectangle (2,1) node[midway] {\large Preamble};
\draw[fill=blue!30] (2,0) rectangle (6,1) node[midway] {\large Access Address};
\draw[] (6,0) rectangle (13,1) node[midway] {\large PDU, max 257 bytes};
\draw[] (13,0) rectangle (15.5,1) node[midway] {\large CRC};

\draw[fill=blue!30] (6,0) rectangle ({6 + 0.735},1);

\draw[thick, <->] (0,1.5) -- (2,1.5) node[midway, above] {\large 1 or 2 bytes};
\draw[thick, <->] (2,1.5) -- (6,1.5) node[midway, above] {\large 4 bytes};
\draw[thick, <->] (6,1.5) -- (13,1.5) node[midway, above] {\large 2 to 257 bytes};
\draw[thick, <->] (13,1.5) -- (15.5,1.5) node[midway, above] {\large 3 bytes};

\node at (7.5,-0.5) {\large No FEC (uncoded)};
\end{tikzpicture}%
}

%% file: coded-phy.tex
    \resizebox{0.95\linewidth}{!}{%
        \begin{tikzpicture}
            \draw (0,0) rectangle (2,1) node[midway] {\large Preamble};
            \draw (2,0) rectangle (6,1) node[midway] {\large Access Address};
            \draw (6,0) rectangle (7,1) node[midway] {\large CI};
            \draw (7,0) rectangle (8.5,1) node[midway] {\large TERM1};
            \draw (8.5,0) rectangle (12.5,1) node[midway] {\large PDU, N bytes};
            \draw (12.5,0) rectangle (14,1) node[midway] {\large CRC};
            \draw (14,0) rectangle (15.5,1) node[midway] {\large TERM2};
    
            \draw[thick, <->] (0,-0.5) -- (8.5,-0.5) node[midway, below] {\large FEC block 1};
            \draw[thick, <->] (8.5,-0.5) -- (15.5,-0.5) node[midway, below] {\large FEC block 2};
    
            \draw[thick, <->] (0,1.5) -- (2,1.5) node[midway, above] {\large 80 $\mu$s};
            \draw[thick, <->] (2,1.5) -- (6,1.5) node[midway, above] {\large 256 $\mu$s};
            \draw[thick, <->] (6,1.5) -- (7,1.5) node[midway, above] {\large 16 $\mu$s};
            \draw[thick, <->] (7,1.5) -- (8.5,1.5) node[midway, above] {\large 24 $\mu$s};
            \draw[thick, <->] (8.5,1.5) -- (12.5,1.5) node[midway, above] {\large N*8*S $\mu$s};
            \draw[thick, <->] (12.5,1.5) -- (14,1.5) node[midway, above] {\large 24*S $\mu$s};
            \draw[thick, <->] (14,1.5) -- (15.5,1.5) node[midway, above] {\large 3*S $\mu$s};
    
            \draw[thick, <->] (0,2.5) -- (8.5,2.5) node[midway, above] {\large S=8 coding};
            \draw[thick, <->] (8.5,2.5) -- (15.5,2.5) node[midway, above] {\large S=2 or S=8 coding};
        \end{tikzpicture}%
    }

%% file: gnu-radio.tex
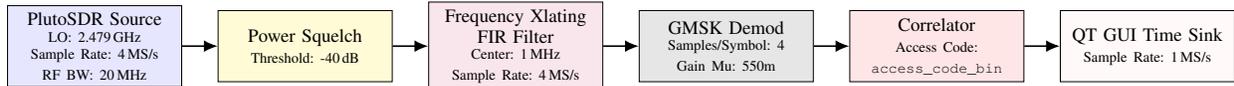
\begin{figure*}[h!]
    \centering
    \resizebox{0.9\linewidth}{!}{%
    \begin{tikzpicture}[node distance=1.6cm and 0.6cm]

\node[block, fill=blue!10] (pluto) {PlutoSDR Source\\ \scriptsize LO: 2.479\,GHz\\ Sample Rate: 4\,MS/s\\ RF BW: 20\,MHz};
\node[block, fill=yellow!20, right=of pluto] (squelch) {Power Squelch\\ \scriptsize Threshold: -40\,dB};
\node[block, fill=purple!10, right=of squelch] (fir) {Frequency Xlating FIR Filter\\ \scriptsize Center: 1\,MHz\\ Sample Rate: 4\,MS/s};
\node[block, fill=gray!20, right=of fir] (demod) {GMSK Demod\\ \scriptsize Samples/Symbol: 4\\ Gain Mu: 550m};
\node[block, fill=red!10, right=of demod] (correlator) {Correlator\\ \scriptsize Access Code: \texttt{access\_code\_bin}};
\node[block, fill=pink!10, right=of correlator] (sink) {QT GUI Time Sink\\ \scriptsize Sample Rate: 1\,MS/s};

\draw[arrow] (pluto) -- (squelch);
\draw[arrow] (squelch) -- (fir);
\draw[arrow] (fir) -- (demod);
\draw[arrow] (demod) -- (correlator);
\draw[arrow] (correlator) -- (sink);

\end{tikzpicture}}
    \caption{Block diagram of the receiver functionality as implemented in \acrlong{gr}.}%
    \label{fig:block-sdr}
\end{figure*}

%% file: circ-mean.tex
\begin{figure}[h!]
    \centering
    \begin{tikzpicture}
  \begin{polaraxis}[
    width=0.8\linewidth,
    grid=both,
    minor grid style={gray!30},
    major grid style={gray!50},
    xlabel={Angle (\unit{\degree})},
    ylabel={Correlation},
    yticklabel style={
      yshift=0.2cm,
    },
    ylabel style={
      yshift=-0.1cm,
      font=\small,     %
    },
    ymin=0,
    ymax=255,
    xmin=-90, xmax=90,           %
    xtick={-90,-45,0,45,90},     %
    anchor=origin, %
    rotate around={90:(current axis.origin)}, %
    clip=false,
    ]
    \addplot[
      mark=*, 
      dashed,
    ] 
    table[
      x=angle,
      y=correlation,
      col sep=comma
    ] {measurments.txt};
  \end{polaraxis}
\end{tikzpicture}
    \caption{Measured correlation when the target is at \SI{0}{\degree} and \SI{50}{\meter} away.}
    \label{fig:circ-meas}
\end{figure}
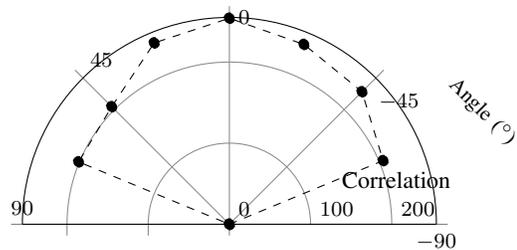